\documentstyle[12pt]{article}
\begin{document}
\title{Heavy Mesons Spectroscopy}
\author
{N. Tazimi \thanks{ nt$_{-}$physics @ yahoo.com}\\ M. Monemzadeh\thanks{monem@kashanu.ac.ir}\\
M. R. Hadizadeh\thanks{hadizade@ift.unesp.br}\\
\it\small $^{1,2}${{Department of Physics, University of Kashan, Iran.}}\\
\it\small$^{3}${{Instituto de Fısica Teorica, UNESP, 01405-900 Sao Paulo, Brazil}}}
\date{}
\maketitle
\begin{abstract}
In this paper, we use Martin and Coulomb-Linear potentials and solve Lippman-Schwinger equation and then identify $b\bar{c}$ energy levels. Moreover, we predict results for such energy levels as that of $t\bar{t}$ (in its short half-life) which is not observed. We showed our results are consistent with previous findings in literature. Also investigating spectrum of eigen-values, we obtain stability interval for Yukawa-Linear potential.
\end{abstract}
\textbf{keyword:}\\
Schwinger equation, Mass spectrum, Strong potential models, Cornel potential, heavy quark \newpage 
\section{Introduction}
Hadrons consist of quarks and antiquarks and follow the principles of Quantum Chorodynamics (QCD). Perturbative measuring of physical properties of hadrons does not always yield consistent results except for the deep inelastic region in which the coupling constant is influenced by asymptotic freedom and is weakened. Therefore, such theories as lattice gauge theory have emerged to consider non-perturbative effects. Among hadrons, mesons are favorable means of exploring strong interaction in the non-perturbative method with strong coupling. Hence, studying meson spectrum could highly help conceive of strong interaction dynamics.\\

Recognizing mass spectrum via solving non-relativistic Schr\"{o}dinger equation with a variety of potentials is a topical issue in different branches of physics \cite{1,2}.  It is worth mentioning that we focus on heavy quark mesons in solving Schr\"{o}dinger equation in order to avoid relativistic problem. Potential models describe the mass spectrum of heavy quarkonium. In this system, the interaction potential is often a confining one. \\

Heavy quarkonium spectrum is a suitable instrument for exploring the static interaction in the infrared region. It also provides an excellent context to study the effects of fine and hyperfine structures. Additionally, solution of Schr\"{o}dinger equation is a major step toward constructing a lot of potential models of quarkonium. In other words, the energy spectrum of the Schr\"{o}dinger through various potentials is obtained in different ways; for example via numerical and variational methods. Hence, many potential models of quarkonium (such as Martin and Coulomb-Linear potential) are used that fit the mass spectrum of $\Psi$ and $\Upsilon$ \cite{3,4,5,6,7}.\\
In Sec.2 we explain our method and solve Schr\"{o}dinger equation for heavy-quark mesons bound state. In Sec.3 introducing $B_{c}$ system, we obtain $B_{c}$ energy levels. Sec.4 is devoted to the estimating of the energy levels of $t\bar{t}$, and in Sec.5 we present stability intervals for $\Psi$ and $\Upsilon$ for Yukawa plus linear potential.\\
\section{Lippmann-Schwinger Equation } 
Since the binding energies in heavy-quark mesons are small compared to their constituent quark masses, one should use relativistic equations to analyze these systems. As we know, relativistic equations are too complicated. Hence, heavy-quark mesons (which have high quark masses compared to their energy levels) are preferred to be used in the present study. Furthermore, Schrodinger equation, the integral form of which is Lippman-Schwinger equation, is employed for the calculations. \\
We can use homogeneous Lippman-Schwinger equation for two particles. Schr\"{o}dinger equation will be:
\begin{equation} 
\mid\psi_{b}>=G_{0}V\mid\psi_{b}> 
\label{1}
\end {equation}
Equation (\ref{1}) in configuration space turns out as:\\
\begin{equation} 
\psi_{b}(r)=-m\sqrt{\pi/2}{\int_{0}^{\infty}}{ dr^\prime} {r^\prime}^{2}{\int_{-1}^{1}} dx^\prime {\int_0}^{2\pi} d\phi ^\prime \frac{exp(-\sqrt{m\vert E_{b}\vert} \vert{r-r^\prime}\vert)}{\vert{r-r^\prime}\vert} V(r^\prime){\psi_{b}(r^\prime)} 
\label{2}
\end {equation}
where $E_{b}$ denotes the binding energy of the system. Equation (\ref{2}) could be rewritten as:
\begin{equation}
{\psi_{b}(r)}=\int_{0}^ {\infty}dr^{'} \int_{-1}^ {1} dx^{'} M(r,r^{'}, x^{'}) {\psi_{b}(r^{'})} 
\label{3}
\end{equation}
where:
\begin{equation} 
M(r,r^{'},x^{'})=-m\sqrt{2\pi}2\pi\frac{exp((-\sqrt{m\vert E_{b}\vert) }\sqrt{r^{2}+\acute{r}^{2}-2r \acute{r} \acute{x}})}{\sqrt{r^{2}+\acute{r}^{2}-2r\acute{r} \acute{x}}}{} \\ 
{ V(\acute{r}^{2})} 
\label{4}
\end{equation}
The eigen-value form of Equation (\ref{3}) is:
\begin{equation}
K(E_{b})\vert\psi_{b}>=\lambda(E_{b})\vert\psi_{b}>
\label{5}
\end{equation}
where $\lambda=1$ is the positive eigen-value of the highest value to solve this equation. We use Gauss-Legendre method and integral equation is transformed into \cite{8}:
\begin{equation}
\psi_{b}(r)=-m\sqrt{2\pi}2\pi \sum_{j=1}^{N_r^\prime} \sum_{i=1}^{N_r^\prime}{W_{r_{i}}^\prime} {W_{x_{j}}^\prime} {r^\prime_{i}}^{2}\frac{exp(-\sqrt{m\vert E_{b}\vert} \rho(r,r^\prime_{i},x^\prime_{j} ))}{\rho(r,r^\prime_{i},x^\prime_{j} )} V(r_{i}^\prime){\psi_{b}(r^\prime_{i})}
\label{6}
\end{equation}
Equation (\ref{6}) could be rewritten as:
\begin{equation}
{\psi_{b}(r)}=\sum_{i=1}^{N_{r^{'}}} N(r,r^{'}_{i}) {\psi_{b}(r^{'}_{i})}
\label{7}
\end{equation}
that:
\begin{equation}
N(r,r^\prime_{i})=-m\sqrt{2\pi}2\pi\sum_{j=1}^{N_r^\prime} {W_{r_{i}}^\prime} {W_{x_{j}}^\prime} {r^\prime_{i}}^{2}\frac{exp(-\sqrt{m\vert E_{b}\vert} \rho(r,r^\prime_{i},x^\prime_{j} ))}{\rho(r,r^\prime_{i},x^\prime_{j} )} V(r_{i}^\prime)
\label{8}
\end{equation}
Then we use a Fortran code that renders the kernel in  Lippman-Schwinger equation diagonal and presents the eigen-values as its output. Appearance of $\lambda=1$ in the output indicates that the energy supposed for the system is the binding energy. This way, we obtained the energy levels for $t\bar{t}$ and $b\bar{c}$.\\
\section{$B_{c}$ Energy Levels} 
Investigation of the spectrum of $B_{c}$ system is a big step toward better calculation of the quantitative properties of quark models and improving QCD sum rules. The $B_{c}$ quarkonium states are capable of submitting an excellent representation of heavy quark dynamics.\\
The $B_{c}$ meson is fascinating in that it enables us to employ phenomenological information derived from the investigation of Bottomonium and Charmonium. \\
This meson is the only heavy meson which is made up of two heavy quarks with different flavors of $b$ and $c$. Charm and Beauty are regarded as heavy mesons because their masses highly exceed the scale of interaction of QCD ($ m_{Q}\gg\Lambda_{QCD}$) \cite{9}. The mass spectrum of this meson lies somewhere between those of $\psi(c \bar{c})$ and $\Upsilon(b \bar{b})$ systems. Therefore $B_{c}$ meson is an interesting topic for study. $B_{c}$ system has been studied experimentally \cite{10} and elaborated on theoretically \cite{11,12,13,14,15,16}.\\
In order to calculate the energy levels of the $B_{c}$ system, we use the Martin potential with these data \cite{6}:\vspace*{0.5cm}
\begin{equation}
\begin{array}{l}
V(r)=A(r \Lambda)^{\nu}+B\\ 
m_{c}=1.65 \quad GeV\\ 
m_{b}=5.04 \quad  GeV\\
A=8.068 \quad  GeV \\
\Lambda =1 \quad  GeV\\
B=6.869 \quad  GeV\\
\nu=0.1 \\
\end{array}
\label{9}
\end{equation} \vspace*{0.5cm}\\
We seek $\lambda=1$ through the eigen-value spectrum in the output. The input energy introduced into the program that creates $\lambda=1$ is the binding energy of each level. Our findings in table(1) show the energy levels of the $B_{c}$ system. In this table, the relativistic corrections are ignored. The results derived from our procedure are consistent with \cite{17} and \cite{18}.\\
Table (\ref{1}) shows that the results with $nS$ states are more consistent with \cite{17,18} than the results with $nP$ and $nD$ states. \newpage 
\begin {table}
\centering
\caption { $B_{c}$ Energy Levels }
\begin{tabular}{|c |c c c|}
\hline$\ state $ &{$\quad 1S $}&{$\quad 2S$}&{$\quad 3S$}\\\hline
ref\cite{17} & {$\quad 6.301 $}&{$ \quad 6.893 $} &{$ \quad 7.237 $} \\
ref \cite{18} &{$\quad 6.344 $}& {$\quad 6.910 $}&{$\quad 7.024 $} \\
\quad this method & {$\quad 6.28 $}&{$ \quad 6.93 $} &{$ \quad 7.12 $} \\ \hline
$\ state $ &{$\quad 2P $} &{$\quad 3P $} & {$\quad 4P$}\\ \hline
ref\cite{17} & {$\quad 6.728 $}& {$ \quad 7.122 $} &{$ \quad 7.395 $} \\
ref$\cite{18} $ &{$\quad 6.763 $}& {$\quad 7.160 $} &{$\quad - $} \\
this method &{$\quad 6.74 $} & {$\quad 7.2 $} & {$\quad7.29 $} \\ \hline
$\ state $ &{$\quad 3D $} & {$\quad 4D $} & {$\quad 5D $}\\ \hline
ref\cite{17} & {$\quad 7.008 $}& {$ \quad 7.308 $} &{$ \quad 7.532 $} \\
ref\cite{18} &{$\quad 7.030 $}& {$\quad 7.365 $}&{$\quad - $} \\
this method &{$\quad 7.19$} & {$\quad 7.25 $} & {$\quad7.34 $} \\ \hline
\end{tabular} 
\end{table}
\section{Predicting Topomonium Energy Levels} 
A number of factors involving hyperfine splitting, spin orbit, and revising can result in better modeling of $\psi(c\bar{c})$ and $\Upsilon(b\bar{b})$. Of course, revising different models is not necessary for topomonium since we want to obtain the approximate position of energy in topomonium spectrum \cite{19}.\\
According to phenomenological studies, t-quark has a short half-life \cite{20}. We consider its mass spectrum in this short half-life.\\
In the present work we employ Coulomb-Linear potential with this form \cite{4,5}:
\begin{equation}
\begin{array}{l}
V(r) = -\frac{A}{r}+Br+C\\
C= -0.542 \quad  GeV \\ 
B=0.185 \quad  GeV^{2} \\ 
A=0.47 \\
m_{t}=21 \quad GeV \cite{19}
\end {array}
\label{10}
\end{equation}
Cornel potential consists of two terms. One is Coulomb potential, which applies to short distances. This term displays the exchange of one gluon. The second term applying to large distances is linear potential that is indicative of quark confinement. We solve  Lippman-Schwinger equation to detect the eigen-value of $\lambda=1$. We estimate the energy levels of topomonium in table (\ref{2}).\newpage 
\vspace*{1cm}
\begin {table}
\centering
\caption { mass spectrum of topomonium with Coulomb-Linear potential coefficient $m_{t}=21  GeV, A=0.47, B=0.185  GeV^{2}, C=-0.542  GeV$}
\begin{tabular}{|c |c c c c c c c|}
\hline $\ state $ & {$ 1S $} &{$ 2S $} & {$ 3S $}& {$ 4S $}& {$ 5S $} & {$ 6S $} & {$ 7S $}\\ \hline
ref \cite{19} & {$\ 40.352 $}&{$ 41.372 $ } &{$41.701$}& {$41.924$}&{$42.108 $} &{$ 42.269 $} &{$ 42.407 $} \\
\quad this method & {$40.1 $}&{$ 40.45 $} &{$ 41.9 $}& {$ 41.98 $}&{$ 42.2 $} &{$ 42.39 $}& {$ 42.8 $} \\ \hline
$\ state $ & \quad & {$ 2P $}& {$ 3P $} & {$4P$}& {$ 5P $}& {$ 6P $} & {$ 7P $} \\ \hline
ref \cite{19} &\quad & {$ 41.33 $} &{$ 41.699 $} & {$ 41.894 $}& {$42.075 $} &{$42.233 $} & {$ 42.376 $} \\
this method &\quad &{$41.02 $} & {$ 41.39 $} & {$42.01 $} & {$42.19 $}& {$ 42.73 $} &{$ 42.85 $} \\ \hline
state &\quad &\quad &{$ 3D $} & {$ 4D $} &{$ 5D $}&{$ 6D $} & {$7D $}\\ \hline
ref \cite{19} &\quad &\quad & {$ 41.615 $}&{$ 41.846 $} &{$ 42.029 $}& {$42.188 $} &{$ 42.335 $} \\
this method &\quad &\quad &{$ 41.79 $} & {$ 41.8 $} & {$ 41.10 $}& {$41.5 $} & {$ 42.85$} \\ \hline
\end{tabular}
\end{table}
\vspace*{0.5cm} Our results are consistent the previous findings in the literature \cite{19}. It is worth mentioning that n-th state masses are obtained from the binding energy according to: 
\begin{equation} 
M_{n,l}=2m_{q}+E_{n,l}
\label{11}
\end{equation}\\
As we know spin dependent interaction in heavy systems is small (of the $m^{-2}$ order) and can be ignored. The binding energy can also be used to calculate the spin-coupling constant A which is defined as: \cite{8,21} :
\begin{equation}
A=E \frac{m_{1}m_{2}}{s_{1}.s_{2}}
\label{12}
\end{equation}
\section{Stability Interval} 
Yukawa plus linear potential is indicative of the appropriateness of our adopted method. Yukawa plus linear potential \cite{1} is of this form :\\ 
\begin{equation} 
\begin{array}{l}
V(r)= a_{1}r+ v_{1}\frac{e^{-\mu_{1}r}}{r}+ \upsilon_{2}\frac{e^{-\mu_{2}r}}{r}\\ 
a_{1}= 1  GeV^{2},\hspace{0.25 cm} \nu_{1} =10 ,\hspace{0.25 cm}\mu_{1}=5 GeV,\hspace{0.25 cm} \mu_{2}=1 GeV,\hspace{0.25 cm} \upsilon_{2} =-5
\end{array}
\label{13}
\end{equation}\\
The eigen-values of $\lambda$ are calculated through solving Schwinger equation. We study the range of $r^{'}$ around r-cutoff \cite{8}. (R-cutoff results from introducing an artificial barrier for the potential). The interval in which $\lambda=1$ will be stability interval of this potential. This interval shows the range where all potential-coefficients, masses, and binding energies are identified accurately. We show the intervals for Charmonium and Bottomonium in table (\ref{3}).\\
\begin {table}
\centering
\caption { stability interval of $\Upsilon$ and $\Psi$}
\begin{tabular} {|c| c c c|} 
\hline Spin & {$\quad Meson$} &{$ E_{b} $} & {$\quad \ Stability\ interval\ ($fm$) $} \\ \hline
S=1 & {$\quad c \bar{c} $}&{$ \quad -206.63 $} &{$ \quad[1.2,2],[2.5,5.2] $} \\
S=1 &{$\quad b \bar{b} $}& {$\quad -891.93 $} &{$\quad [1,2],[2, 2.5] $} \\ \hline 
S=0 &{$\quad c\bar{c} $} &{$\quad -619.91$} &{$\quad [1,2.1],[2.2,5.2]$}\\ 
S=0 &{$\quad b \bar{b}$}&{$\quad -573.3 $}&{$ \quad[0.22,2],[2.2,12] $} \\ \hline 
\end{tabular}
\end{table} 
\section{Conclusion}
Through the procedure discussed, we solve the Schwinger equation using Martin and Coulomb-Linear potential for the purpose of solving identify the eigen-value problem of quark and antiquark bound states. With this very method, we managed to the energy levels of $b\bar{c}$ and $t\bar{t}$ heavy mesons. The results comply with those in the literature.\\
Binding energy is important in view of the fact that it could be employed to obtain the n-th state mass and coupling coefficient. Through the analysis of the eigen-value spectrum, we identified the stability intervals in which all the potential coefficients of Yukawa plus linear potential, masses, and binding energies are well identified.\\


\begin {thebibliography}{10}
\bibitem{1} joseph P .Day, Joseph E. McEwen and Zolt\`{a}n Papp, Few Body Syst. \textbf{47}, 17 (2010).
\bibitem{2} J. McEwen, JR. Day, A. GeV. D\textbf{17}, 3090 (1987).
\bibitem{3} C. Quigg, J.L.Rosner, phonzalez, Z. Papp and W. Plessas, Few Body Syst.\textbf{47} , 227 (2010).
\bibitem{4} E. Eichten et al., phys. Rev. D\textbf{17}, 3090 (1978).
\bibitem{5}  E. Eichten et al., phys.Rev. D\textbf{21}, (1976).
\bibitem{6} A. Martin, Phys. Lett. 93B,338 (1980); Phys. Lett. 100B,\textbf{511} (1981).
\bibitem{7} H. Quigg, Crater,P, Van Alstine,Phys. Lett. 100B,\textbf{166} (1981).
\bibitem{8} M. Monemzadeh and M.Hadizade anh N.Tazimi, Int. J. of theoretical Phys. Vol 50 No. \textbf{3} (2011).
\bibitem{9} V.V.Kiselev [arXiv:hep-ph/0805.4329]. 
\bibitem{10} Hikasa K et al, PDG. Phys.Rev.D \textbf{45}(II) S1 (1992).
\bibitem{11} S. Godfrey, Phys. Rev. D\textbf{70} (2004) 054017.
\bibitem{12} S.M. Ikhdair and R. Sever, Int. J. Mod. Phys. A\textbf{18} (2003) 4215.
\bibitem{13} S.M. Ikhdair and R. Sever, Int. J. Mod. Phys. A\textbf{19} (2004) 1771; ibid. A\textbf{20} (2005) 4035. [arXiv:hep-ph/0406005].
\bibitem{14} M. Baldicchi and G.M. Prosperi, Phys. Lett. B\textbf{436} (1998) 145; ibid. Phys. Rev. D \textbf{62}.
(2000) 114024; ibid. Fiz. B \textbf{8} (1999) 251.
\bibitem{15} D. Ebert, R.N. Faustov and V.O. Galkin, Phys. Rev. D\textbf{67} (2003) 014027.
\bibitem{16} D. Ebert, R. N. Faustov, V. O. Galkin, Mod. Phys. Lett. A\textbf{17} (2002) 803; [arXiv:hep-ph/0210381].
\bibitem{17} S.S.Gershtein et al., Yad.Fiz.\textbf{48}(1988) 515 [Sov.J.Nucl.Phys. \textbf{48} (1988). 326];S.S.Gershtein, A.K.Likhoded, S.R.Slabospitsky, Int.J.Mod.Phys. A\textbf{6}(13) (1991) 2309.
\bibitem{18} Y.-Q.Chen, Y.-P.Kuang, Phys.Rev. D\textbf{46} (1992) 1165.
\bibitem{19} P.Aretymowicz, Acta physica Polonica , B\textbf{15}(1984).
\bibitem{20}Andre H. Hoang, Top Threshold Physics. International Workshop on Top Quark Physics, Coimbra, Portugal (2006).
\bibitem{21} F. Halzen , A.D. Martin, "Quarks and Leptons", (New York: Wiley 1984).
\end{thebibliography}
\end{document}